\documentclass[reprent,aps,prb,twocolumn,
superscriptaddress,nofootinbib]{revtex4-2}
\usepackage{amsmath,amssymb}
\usepackage{graphicx}
\usepackage{bm}
\usepackage{multirow}
\usepackage{hyperref}
\usepackage{braket}

\usepackage{color}

\allowdisplaybreaks[2]

\begin{document}

\title{
Analysis of photo-induced chirality and magnetic toroidal moment \\ based on Floquet formalism
}
\author{Satoru Hayami}
\email{hayami@phys.sci.hokudai.ac.jp}
\affiliation{Graduate School of Science, Hokkaido University, Sapporo 060-0810, Japan}
\author{Ryota Yambe}
\affiliation{Department of Applied Physics, The University of Tokyo, Tokyo 113-8656, Japan }
\author{Hiroaki Kusunose}
\affiliation{Department of Physics, Meiji University, Kawasaki 214-8571, Japan}
\affiliation{Quantum Research Center for Chirality, Institute for Molecular Science, Okazaki 444-8585,}

\begin{abstract}
We analyze the condition of photo-induced atomic-scale chirality and magnetic toroidal moment. 
By performing a high-frequency expansion in the Floquet formalism, we derive an effective static model Hamiltonian from the spinful $s$-$p$  hybridized model of a single atom interacting with an electromagnetic wave with a particular polarization. 
The lowest-order and third-order contributions in the high-frequency expansion give rise to the coupling to induce an electric toroidal monopole corresponding to microscopic chirality, while the second-order contribution provides the coupling to induce a magnetic toroidal dipole.
We also discuss the condition of the polarization of the electromagnetic wave and induced multipoles.
Our results stimulate a new direction of controlling unconventional multipoles by electromagnetic waves.
\end{abstract}

\maketitle

\section{Introduction}

The electric and magnetic dipolar moments are fundamental resources in electromagnetism, and they are well controlled by their conjugate electric and magnetic fields.
It has been extensively developed a much richer landscape of multipolar moments both in real and momentum spaces, originating from the interplay between electron's charge, spin, orbital, and its underlying lattice structure.
The resulting unconventional electronic ordering of multipoles can have unique properties that make them appealing not only from a fundamental quantum many-body perspective, but also for new devices and applications.
However, it is often difficult to control the multipolar moments as their conjugate fields are not provided by a simple static field.

Among a variety of electronic multipoles~\cite{Hayami_PhysRevB.98.165110, Yatsushiro_PhysRevB.104.054412, kusunose2022generalization, Kusunose_PhysRevB.107.195118}, a magnetic-toroidal (MT) dipole (denoted by a symbol $\bm{T}$), that is time-reversal odd polar vector~\cite{dubovik1975multipole, dubovik1986axial, dubovik1990toroid, Spaldin_0953-8984-20-43-434203, kopaev2009toroidal},  becomes an origin of an asymmetric band modulation~\cite{Yanase_JPSJ.83.014703, Hayami_doi:10.7566/JPSJ.84.064717, Hayami_doi:10.7566/JPSJ.85.053705}, linear magnetoelectric effect~\cite{EdererPhysRevB.76.214404, Spaldin_0953-8984-20-43-434203, Hayami_PhysRevB.90.081115, hayami2016emergent,thole2018magnetoelectric, Shitade_PhysRevB.98.020407,Gao_PhysRevB.97.134423, thole2020concepts}, nonlinear charge/spin transport~\cite{Watanabe_PhysRevResearch.2.043081, Yatsushiro_PhysRevB.105.155157, Hayami_PhysRevB.106.024405, Wang_PhysRevLett.127.277201, Liu_PhysRevLett.127.277202, Kirikoshi_PhysRevB.107.155109}, and so on.
Meanwhile, an electric-toroidal (ET) dipole (denoted by $\bm{G}$), time-reversal even axial vector~\cite{Hlinka_PhysRevLett.113.165502, Hlinka_PhysRevLett.116.177602, Hayami_doi:10.7566/JPSJ.91.113702}, is a microscopic variable of the so-called ferroaxial ordering~\cite{jin2020observation, hayashida2020visualization, cheong2022linking}.
Moreover, an ET monopole (denoted by $G_{0}$), time-reversal even pseudoscalar, has been identified as a microscopic representation of chirality, whose sign corresponds to each handedness~\cite{kusunose2020complete, Oiwa_PhysRevLett.129.116401,kishine2022definition}.

\begin{figure}[t!]
\begin{center}
\includegraphics[width=1.0\hsize]{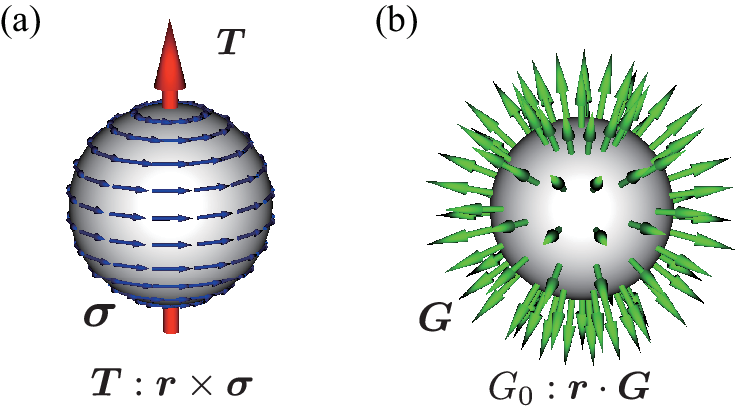} 
\caption{
\label{fig:ponti}
(a) Magnetic toroidal (MT) dipole $\bm{T}$ (the red arrow) having the same symmetry of the outer product of the position $\bm{r}$ and spin $\bm{\sigma}$ (the blue arrows).
(b) Electric toroidal (ET) monopole $G_0$, the same symmetry of the scalar product of $\bm{r}$ and the ET dipole $\bm{G}$ (the green arrows). 
}
\end{center}
\end{figure}

These multipoles are mutually converted with each other, when one recognizes the following relations from a symmetry point of view, (i) a composite object $G_{0}\leftrightarrow\bm{\sigma}\cdot\bm{T}$ with spin $\bm{\sigma}$, (ii) cluster objects $G_{0}\leftrightarrow\bm{r}\cdot\bm{G}$ and $\bm{T}\leftrightarrow\bm{r}\times\bm{\sigma}$ at a position $\bm{r}$, and so on.
In other words, the chirality is a source or sink of $\bm{G}$ in real space, and that of $\bm{T}$ in spin space.
The schematic pictures of such cluster objects are shown in Figs.~\ref{fig:ponti}(a) and \ref{fig:ponti}(b).
Such an ET monopole has been discussed as an unconventional electronic order parameter, such as in the elemental tellurium~\cite{Oiwa_PhysRevLett.129.116401}, Cd$_2$Re$_2$O$_7$~\cite{Hayami_PhysRevLett.122.147602}, and URu$_2$Si$_2$~\cite{hayami2023chiral}.

Static conjugate field of such less familiar but fundamental multipoles is usually nonuniform, for instance, $\bm{\nabla}\times \bm{B}$ is a conjugate field for $\bm{T}$, $\bm{\nabla}\times \bm{E}$ is for $\bm{G}$, and $\bm{E}\cdot(\bm{\nabla}\times\bm{E})+\bm{B}\cdot(\bm{\nabla}\times\bm{B})$ is for $G_{0}$, which is known as an optical chirality, Zilch~\cite{Tang_PhysRevLett.104.163901}.
Unfortunately, $\bm{\nabla}\times \bm{B}$ and $\bm{\nabla}\times \bm{E}$ are difficult to generate experimentally in general, however, they are related to the time derivative of electromagnetic fields through Maxwell equations.
Therefore, an electromagnetic wave can be a useful conjugate field for the toroidal type of electronic multipoles.

In the present study, we explore effective interactions between the toroidal multipolar moments and an electromagnetic wave, by adopting the Floquet formalism that converts a periodic time-dependent Hamiltonian into a time-independent one with a couple of replicas of the original nonperturbative Hamiltonian with a constant energy difference of the applied frequency~\cite{Eckardt_RevModPhys.89.011004, oka2019floquet, rudner2020band}.
Based on a prototype of spinful $s$-$p$ hybridized model of a single atom containing all the relevant multipoles, we find the effective interaction between the toroidal multipoles and an electromagnetic wave with a particular polarization~\cite{kishine2022definition}.
Our results suggest a way to induce a variety of multipoles by electromagnetic fields, which leads to a further understanding of fascinating multipolar phenomena in condensed matter. 

The paper is organized as follows. 
In Sec.~\ref{sec: Model}, we introduce the prototype model Hamiltonian and a bare coupling to an electromagnetic wave. 
Then, we show the effective Hamiltonian based on the Floquet formalism in Sec.~\ref{sec: Effective Hamiltonian}, where the material chirality and MT multipolar moments are induced by the third-order and second-order contributions, respectively.
Section~\ref{sec: summary} summarizes the paper.
In Appendix~\ref{sec: Appendix1}, we show the definition of multipoles in the present $s$-$p$ model.
Then, we derive the bare interaction between a matter and an electromagnetic wave in Appendix~\ref{sec:coupling}.
We discuss the effect of the additional bare interaction in Appendix~\ref{sec: Appendix2}.

\section{Model}
\label{sec: Model}

In this section, we first introduce the prototype of a spinful $s$-$p$ hybridized model of a single atom. 
We show that the model has all of the relevant multipoles in question in Sec.~\ref{sec: Static Hamiltonian}.
Then, we explain the bare interaction between the above atomic system and an electromagnetic wave with linear or circular polarization in Sec.~\ref{sec: Dynamical Hamiltonian}.

\subsection{Single-atom Hamiltonian}
\label{sec: Static Hamiltonian}

Let us consider a spinful $s$-$p$ orbital model with one $s$ orbital and three $p$ orbitals.
The Hamiltonian is given by 
\begin{align}
H_{0}=\sum_{\alpha\sigma}\Delta_{\alpha}c^{\dagger}_{\alpha\sigma}c_{\alpha\sigma}^{}
+ \frac{\lambda}{2} \sum_{\alpha\sigma,\alpha'\sigma'}(\bm{l} \cdot \bm{\sigma})^{\alpha\alpha'}_{\sigma\sigma'}c^{\dagger}_{\alpha\sigma}c_{\alpha'\sigma'}^{},
\end{align}
where $c^{\dagger}_{\alpha\sigma}$ and $c_{\alpha\sigma}$ are creation and annihilation operators of the orbital state $\alpha=s$, $p_{x}$, $p_{y}$, $p_{z}$ and spin state $\sigma$.
$\Delta_{s}$ and $\Delta_{p_{x}}=\Delta_{p_{y}}=\Delta_{p_{z}}$ are the energy levels of $s$ and $p$ orbitals, respectively, and we set $\Delta=\Delta_{p_{z}}-\Delta_{s}$.
The second term represents the atomic spin-orbit coupling (SOC) where $\bm{l}$ and $\bm{\sigma}$ are the dimensionless orbital angular momentum and the vector of Pauli matrices.

In this Hilbert space of $(1+3)\times2=8$ states, there are $8\times8=64$ multipole degrees of freedom in total.
Without the spin degrees of freedom, we have $4\times4=16$ multipoles; electric monopoles $Q_{0s}$, $Q_{0p}$, electric dipoles $\bm{Q}=(Q_x, Q_y, Q_z)$, electric quadrupoles $Q_u$, $Q_v$, $Q_{yz}$, $Q_{zx}$, $Q_{xy}$, magnetic dipoles $\bm{M}=(M_x, M_y, M_z)$, and MT dipoles $\bm{T}=(T_x, T_y, T_z)$.
The other 48 multipoles in spinful space are obtained by the direct products of the above spinless multipoles and $\bm{\sigma}$, which are summarized in Appendix~\ref{sec: Appendix1}.
It should be emphasized that the model contains the ET monopole, $G_{0}^{(s)}$, ET dipole, $\bm{G}^{(s)}$, MT dipole, $\bm{T}^{(s)}$, and the anisotropic dipole, $\bm{M}_{a}^{(s)}$, where the superscript (s) represents a spin-dependent multipole.

\subsection{Interaction between matter and electromagnetic wave}
\label{sec: Dynamical Hamiltonian}

We consider the interaction between the $s$-$p$ hybridized model and an electromagnetic wave as
\begin{align}
\label{eq:Ht}
H_{\rm int}(t) =  -\bm{\mu}_{\rm e}\cdot\bm{E}(t) -\bm{\mu}_{\rm m}\cdot\bm{B}(t),
\end{align}
where $\bm{E}(t)$ and $\bm{B}(t)$ represent the electric and magnetic fields, and $\bm{\mu}_{\rm e}=-e\bm{r}+\bm{\mu}_{\rm e}'$ and $\bm{\mu}_{\rm m}=-\mu_{\rm B}(\bm{l}+\bm{\sigma})+\bm{\mu}_{\rm m}'$ are the electric and magnetic dipoles, respectively.
The derivation of this interaction is given in Appendix~\ref{sec:coupling}.
$\bm{\mu}_{\rm e}'$ and $\bm{\mu}_{\rm m}'$ are the spin-orbit corrections, and we omit them hereafter.
The contributions of these additional terms are discussed in Appendix~\ref{sec: Appendix2}.
In the $s$-$p$ orbital space, $\braket{s,p|\bm{\mu}_{\rm e}|s,p}=-e\braket{r}_{sp}\bm{Q}$, and $\braket{s,p|\bm{\mu}_{\rm m}|s,p}=-\mu_{\rm B}(\bm{M}+\bm{M}_{s}^{\rm (s)}+\bm{M}_{p}^{\rm (s)})$ where $\braket{r}_{sp}$ is the average between $s$ and $p$ radial wave functions.

The electromagnetic wave propagating to $+z$ direction as defined from the point of view of the source is expressed as
\begin{align}
\bm{E}(t)&= E_0(\cos\eta\cos\Omega t,\sin\eta\sin\Omega t ,0), \\
\bm{B}(t)&= B_0(-\sin\eta\sin\Omega t, \cos\eta\cos\Omega t,0), 
\end{align}
where $\Omega$ is the frequency of the electromagnetic wave.
The parameter $\eta$ ($-\pi<\eta \leq \pi$) determines the polarization of $\bm{E}(t)$ component; $\eta=0,\pi$ ($\eta=\pm\pi/2$) is a linear polarization along the $x$ ($y$) axis, while $\eta=\pi/4$ ($-\pi/4$) represents a right (left) circular polarization.
From the Maxwell equation, $E_{0}=B_{0}$ in the Gaussian unit, and in what follows, we use $E=-e\braket{r}_{sp}E_{0}$ and $B=-\mu_{\rm B}B_{0}$.

\section{Effective Hamiltonian}
\label{sec: Effective Hamiltonian}

Since the interaction Hamiltonian $H_{\rm int}(t)$ has a periodicity of $2\pi/\Omega$ in time, the Floquet formalism can be applied~\cite{Eckardt_RevModPhys.89.011004, oka2019floquet, rudner2020band}. 
In the high-frequency region, the total Hamiltonian $H_{0}+H_{\rm int}(t)$ turns into the time-independent effective Hamiltonian by performing the Fourier transformation as $H_0+H_{\rm int}(t) =\sum_m e^{-im\Omega t}H^{(m)}$ with integer $m$ and high-frequency expansion in the Floquet formalism. 
The specific expressions up to $\Omega^{-3}$ are given by~\cite{eckardt2015high, Mikami_PhysRevB.93.144307}
\begin{widetext}
\begin{align}
\label{eq:Ham}
&
H_\mathrm{eff} = H_{0} + H_1 + H_{2} + H_{3} + O(\Omega^{-3}),
\\&\quad
\label{eq:Ham1}
H_{1} = \sum_{m\neq 0}\frac{[H^{(-m)},H^{(m)}]}{ 2m \Omega}
= \frac{[H^{(-1)},H^{(1)}]}{\Omega},
\\&\quad
\label{eq:Ham2}
H_2 = \sum_{m \neq 0} \frac{[[H^{(-m)},H^{(0)}],H^{(m)}]}{2m^2\Omega^2}
=
\frac{1}{2\Omega^{2}}\biggl[[[H^{(-1)},H^{(0)}],H^{(1)}]+[[H^{(1)},H^{(0)}],H^{(-1)}]
\biggr],
\quad
\\&\quad
\label{eq:Ham3}
H_{3} = \sum_{m \neq 0} \frac{[[[H^{(-m)},H^{(0)}],H^{(0)}],H^{(+m)}]}{2m^3\Omega^3}
=
\frac{1}{2\Omega^3}
\biggl[
[[[H^{(-1)},H^{(0)}],H^{(0)}],H^{(1)}]
-[[[H^{(1)},H^{(0)}],H^{(0)}],H^{(-1)}]
\biggr],
\end{align} 
where $[A,B]=AB-BA$ is a commutator. 
It is noted that there is no contributions of $m \neq 0,\pm 1$.
\end{widetext}

In the following, we discuss the effective Hamiltonian of each order $H_{n}$ ($n=1,2,3$) in Secs.~\ref{sec: First-order contribution}-\ref{sec: Third-order contribution}.
After we show the results of each contribution, the numerical evaluation of the effective Hamiltonian is given in Sec.~\ref{sec: Numerical evaluation}, and the symmetry relation in the presence of the electromagnetic wave is discussed in Sec.~\ref{sec: Symmetry analysis}.

\subsection{First-order term}
\label{sec: First-order contribution}

The first-order term is given by 
\begin{align}
\label{eq: Ham1}
H_1=-\frac{\sin 2\eta}{12\Omega} \left[\left(E^2+3 B^2\right) M_z +6B^2   (M^{\rm (s)}_{s,z}+M^{\rm (s)}_{p,z}) \right],  
\end{align}
The electric field induces the orbital magnetization, while the magnetic field induces both the orbital and spin ones. 
As this term is proportional to $\sin 2\eta$, the $z$ component of the magnetization is not induced by the electromagnetic wave with the linear polarization, $\eta=0$, $\pi$ or $\pm\pi/2$, and the magnetizations induced by the circular polarization depend on the sign of the polarization. 
A similar result was obtained for different models in previous studies~\cite{Takayoshi_PhysRevB.90.085150, Takayoshi_PhysRevB.90.214413, Sato_PhysRevLett.117.147202, Topp_PhysRevB.105.195426,higashikawa2018floquet,PhysRevLett.128.037201,Yambe_PhysRevB.108.064420}.

\subsection{Second-order term}
\label{sec: Second-order contribution}

The second-order term is given by
\begin{align}
&
H_2=\frac{1}{4\Omega^2} [(H^{\perp}_{2a}+H^{\parallel}_{2a})
+ \cos 2\eta(H^{\perp}_{2b}+H^{\parallel}_{2b})],
\\&\quad
\label{eq: H2cI}
H^{\perp}_{2a}=EB \left(
 3 \Delta  T_z -\sqrt{2}\lambda T^{\rm (s)}_z
\right),
\\&\quad
H^{\parallel}_{2a}= \frac{E^2 \Delta}{9} \left( 
6  Q_{0s} -2  Q_{0p}+5  Q_u
\right)
\cr&\hspace{1.5cm}
-\frac{\lambda}{6\sqrt{6}} \left(E^2+3 B^2\right) \left(
2\sqrt{2} Q^{\rm (s)}_0 + Q^{\rm (s)}_u
\right),
\\&\quad
\label{eq: H2cII}
\mathcal{H}^{\perp}_{2b}= -\frac{E B \lambda }{ \sqrt{2}}  M^{\rm (s)}_{xy},
\\&\quad
H^{\parallel}_{2b}= -\frac{5 E^2 \Delta }{3 \sqrt{3}} Q_v+\frac{\lambda \left(E^2-3 B^2\right) }{6 \sqrt{2}}Q^{\rm (s)}_v, 
\end{align}
where $H_2$ consists of the part independent of $\eta$, $H_{2a}$, and that proportional to $\cos 2\eta$, $H_{2b}$; the former exists irrespective of the polarization, while the latter vanishes for the electromagnetic wave with the exact circular polarization, $\eta= \pm \pi/4$.
Here, we denote the terms proportional to $EB$ with the superscript, $\perp$, otherwise with $\parallel$.
Note that in the case of the linear polarization, $\eta=0$, $\pi$ or $\pm \pi/2$, the present term becomes the lowest order of the expansion.
It is clear that the SOC is necessary to induce the spin-dependent multipoles denoted by the superscript (s). 

One of the characteristic features in $H_2$ is the emergence of two MT dipoles in Eq.~(\ref{eq: H2cI}). 
This is intuitively understood by the fact that the electric and magnetic fields are orthogonal with each other, i.e., $\bm{E} \times \bm{B}$ is always finite, which has the same symmetry of MT dipole.
It is noted that both $T_z$ and $T^{\rm (s)}_z$ are induced irrespective of the polarization, $\eta$, although they exhibit different model parameter dependence: The former is related to the atomic energy level $\Delta$, while the latter is related to the SOC $\lambda$. 
This means that the MT dipole can be induced by the electromagnetic wave even without the SOC.

Another characteristic feature is the appearance of the magnetic quadrupole $M^{\rm (s)}_{xy}$ in Eq.~(\ref{eq: H2cII}), which becomes a source of the linear transverse magnetoelectric effect and current-induced distortion (magnetopiezoelectric effect)~\cite{Watanabe_PhysRevB.96.064432, Hayami_PhysRevB.104.045117}.  
Since this term is proportional to $\cos 2\eta$, $M^{\rm (s)}_{xy}$ can be induced only by using an elliptic or linear polarization, $\eta \neq \pm \pi/4$.
This is consistent with the symmetry lowering by the elliptic polarization, i.e., it breaks rotational symmetry in the $xy$ plane.
Note that the photo-induced magnetic quadrupole also requires the SOC.

\subsection{Third-order term}
\label{sec: Third-order contribution}

The third-order term is given by 
\begin{align}
&
H_3=\frac{\sin 2\eta}{8\Omega^3} (H^{\perp}_3 + H^{\parallel}_3),
\\&\quad
H^{\perp}_3 = 
-\sqrt{6}E B  \lambda
\left[
 \sqrt{2} (2 \Delta +\lambda)   G^{\rm (s)}_0 +(\Delta-\lambda)   G^{\rm (s)}_u
\right],
\\&\quad
H^{\parallel}_3 = -\frac{1}{3}  \left[2 E^2 \Delta^2+ \left(E^2-3 B^2\right)\lambda^2 \right] M_z
\cr&\hspace{1.2cm}
 +\frac{E^2 \lambda}{3}  (4 \Delta-\lambda)  M^{\rm (s)}_{s,z}
\cr&\hspace{1.2cm}
 -\frac{\lambda }{9}  \left[ 4  E^2 \Delta- \left(E^2-24 B^2\right)\lambda  \right] M^{\rm (s)}_{p,z}
\cr&\hspace{1.2cm}
 + \frac{5\sqrt{10}\lambda }{36}   \left[ 4 E^2 \Delta - \left(E^2+3 B^2\right) \lambda  \right] M^{\rm (s)}_{a,z}.
\end{align}
It is evident that the SOC is also necessary to induce the spin-dependent multipoles.
Similar to $H_1$, the third-order term $H_3$ is also proportional to $\sin 2\eta$, and hence it vanishes when the linear polarization is used.
It should be noted that the ET monopole $G^{\rm (s)}_0$ representing microscopic chirality is induced by the elliptic or circular polarization as shown in $H^{\perp}_3$.
An ET quadrupole $G^{\rm (s)}_{u}$ is also induced;  $G^{\rm (s)}_{u}$ represents the photo-induced monoaxial anisotropy of chirality with respect to the propagation direction even though the atomic Hamiltonian is spherical.
The handedness of chirality can be controlled by the polarization of the electromagnetic wave, that is to say, the sign of the expectation value of $G^{\rm (s)}_0$ and $G^{\rm (s)}_{u}$ is reversed by using the opposite polarization. 

Although it is shown here that the ET monopole is induced in the third order by neglecting the spin-orbit corrections of $\bm{\mu}_{\rm e}$ in Eq.~(\ref{eq:Ht}), it appears in the first order when we consider the spin-orbit correction as discussed in Appendix~\ref{sec: Appendix2}.
It is also noted that the anisotropic magnetic dipole $M^{\rm (s)}_{a,z}$ is induced, although they are not directly coupled to the magnetic field without the spin-orbit correction of $\bm{\mu}_{\rm m}$. 
$M^{\rm (s)}_{a,z}$ plays a significant role in the anomalous Hall effect even without the uniform magnetization~\cite{vsmejkal2020crystal, Hayami_PhysRevB.103.L180407, Chen_PhysRevB.106.024421}, which has been studied extensively in various materials, such as RuO$_2$~\cite{feng2022anomalous, tschirner2023saturation} and the organic conductor~\cite{Naka_PhysRevB.102.075112}.

\subsection{Numerical evaluation}
\label{sec: Numerical evaluation}

\begin{figure}[t!]
\begin{center}
\includegraphics[width=0.8\hsize]{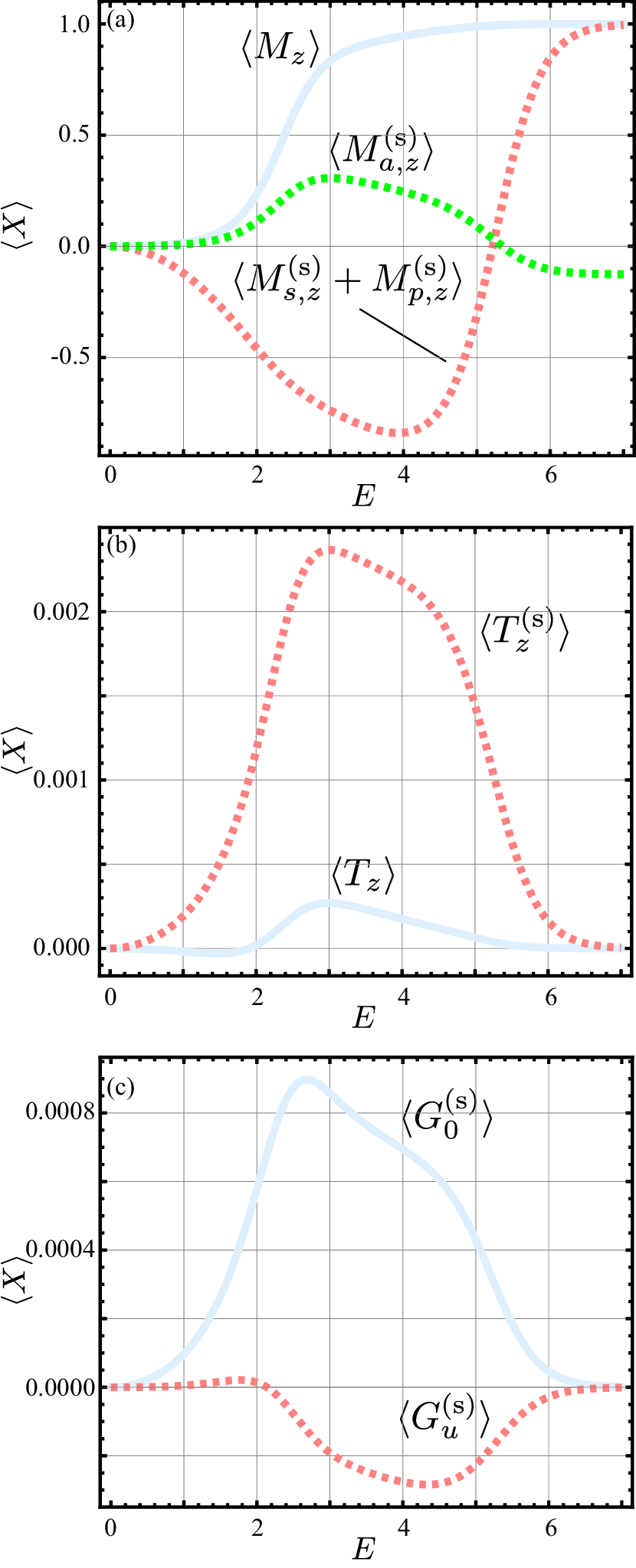} 
\caption{
\label{fig:expec}
The electric-field strength $E$ dependence of the expectation value of multipoles, (a) $\langle M_z \rangle$, $\langle M^{\rm (s)}_{s,z} + M^{\rm (s)}_{p,z} \rangle$, and $\langle M^{\rm (s)}_{a,z} \rangle$, (b) $\langle T_z \rangle$ and $\langle T^{\rm (s)}_z \rangle$, and (c) $\langle G^{\rm (s)}_0 \rangle$ and $\langle G^{\rm (s)}_u \rangle$ at $\Delta=1$, $\lambda=0.7$, $\Omega=2$, $\eta=\pi/4$, $B=E/274$, and $T=0.1$.
}
\end{center}
\end{figure}

To demonstrate the photo-induced multipoles discussed in the previous subsections, we compute the expectation values of each multipole in $H_{\rm eff}$ in Eq.~(\ref{eq:Ham}). 
We set the model parameters as follows: $\Delta=1$, $\lambda=0.7$, $\Omega=2$, $\eta=\pi/4$, and the temperature $T=0.1$; the energy levels for the $s$ orbital with $j=1/2$, the $p$ orbital with $j=1/2$, and the $p$ orbital with $j=3/2$ are located at $-\Delta$, $-\lambda$, and $\lambda/2$, respectively, and $\Delta$ is the energy unit.
We set $B=E/274$ from the ratio of $e a_{\rm B}/\mu_{\rm B}$, where $\langle r \rangle_{sp}$ is regarded as the Bohr radius $a_{\rm B}$ and $E_0 = B_0$.
Figure~\ref{fig:expec} shows the $E$ dependence of the induced strength of the representative multipoles. 
The results for the four magnetic dipoles are shown in Fig.~\ref{fig:expec}(a), where $E^2$ dependence is clearly obtained for small $E$.

Figures~\ref{fig:expec}(b) and \ref{fig:expec}(c) show the behaviors of two MT dipoles, ET monopole, and ET quadrupole, respectively.
All the quantities in Figs.~\ref{fig:expec}(b) and \ref{fig:expec}(c) are induced by $E$, and their magnitudes are enhanced around the two times of energy difference between the $s$ orbital with $j=1/2$ and the $p$ orbital with $j=3/2$.
We confirm that the sign of multipoles in Figs.~\ref{fig:expec}(a) and \ref{fig:expec}(c) is reversed when we apply the electric field with the left circular polarization ($\eta=-\pi/4$).

\subsection{Symmetry analysis}
\label{sec: Symmetry analysis}

\begin{figure}[t!]
\begin{center}
\includegraphics[width=1.0\hsize]{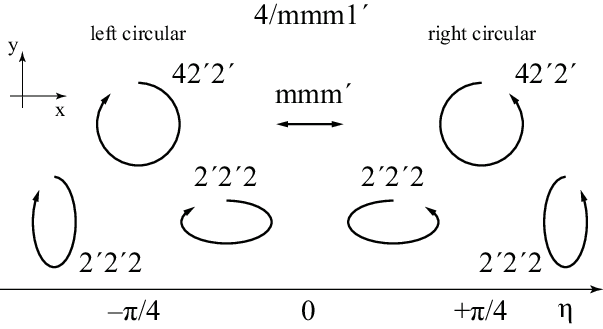} 
\caption{
\label{fig:sym}
Symmetry lowering of the magnetic point group from 4/mmm1$'$ by the electromagnetic wave with the polarization $\eta$.
}
\end{center}
\end{figure}

Finally, let us discuss the relationship between symmetry lowering by applying the electromagnetic wave and photo-induced multipoles. 
We suppose that the symmetry of the static Hamiltonian is the tetragonal $4/mmm1'$ symmetry where the electromagnetic wave propagates along the $z$ direction.
Then, the electric monopole $Q_0$ and the electric quadrupole $Q_u$ belong to the totally symmetric irreducible representation. 
When we consider the circular polarization $\eta= \pm \pi/4$, the magnetic dipole $M_z$, the MT dipole $T_z$, the ET monopole $G_0$, and the ET quadrupole $G_u$ are induced, which indicates the symmetry lowering to $42'2'$~\cite{Yatsushiro_PhysRevB.104.054412, Yambe_PhysRevB.108.064420}.
Meanwhile, when we consider the linear polarization $\eta=0$ or $\pi/2$, the MT dipole $T_z$, magnetic quadrupole $M_{xy}$, and the electric quadrupole $Q_v$ are induced, while the magnetic dipole and the ET monopole are not, indicating that the symmetry is reduced to $mmm'$. 
When we consider the elliptic polarization, all the multipoles mentioned above are induced, which corresponds to $2'2'2$.
We summarize the above classification in Fig.~\ref{fig:sym} and Table~\ref{table: light}. 

It is noteworthy to mention the effective coupling among $M_z$ (time-reversal odd axial vector), $T_z$ (time-reversal odd polar vector), and $G_0$ (time-reversal even pseudoscalar).
Since the product of $M_z T_z G_0$ belongs to the totally symmetric irreducible representation under any point group, one of three quantities is induced once the other two quantities are induced. 
In other words, the microscopic MT dipole and ET monopole are related to each other via the magnetic dipole. 
This implies that the MT dipole and ET monopole are further modulated by applying the additional static magnetic field along the $z$ direction. 

\begin{table}[t!]
\caption{
Classification of magnetic point group (MPG), polarization, and multipole. 
See also Fig.~\ref{fig:sym}.
In the second to fourth lines, the additionally induced multipoles under the electromagnetic wave are presented.
}
\label{table: light}
\footnotesize
\centering{
\begin{tabular}{ccc} \hline 
MPG & Polarization $\eta$ & Multipole \\  \hline
$4/mmm1'$ & -- & $Q_0$, $Q_u$ \\
$42'2'$ & $\pm \pi/4$ &$M_z$, $T_z$, $G_0$, $G_u$  \\
$mmm'$ & $0$, $\pm \pi/2$, $\pi$ & $T_z$, $M_{xy}$, $Q_v$ \\
$2'2'2$ & other & $M_z$, $T_z$, $M_{xy}$, $G_0$, $G_u$, $Q_v$ \\
\hline
\end{tabular}
}
\end{table}

\section{Summary}
\label{sec: summary}

In summary, we have analyzed the condition of the photo-induced unconventional multipoles based on the Floquet formalism.
We have derived the relation between the multipoles and the polarization of the electromagnetic wave in the high-frequency regime.
The MT dipole is induced irrespective of the polarization, while the ET monopole corresponding to the microscopic chirality is induced by the circular (elliptic) polarization. 
In both cases, the spin-orbit coupling is necessary when the induced multipoles are spin-dependent.
Our result will provide another direction for controlling unconventional multipoles by electromagnetic waves.

\begin{acknowledgments}
This research was supported by JSPS KAKENHI Grants Numbers JP21H01037, JP22H04468, JP22H00101, JP22H01183, JP23KJ0557, JP23H04869, JP23K03288, JP23H00091, and by JST PRESTO (JPMJPR20L8) and JST CREST (JPMJCR23O4), and the grants of Special Project (IMS program 23IMS1101), and OML Project (NINS program No. OML012301) by the National Institutes of Natural Sciences.  
R.Y. was supported by Forefront Physics and Mathematics Program to Drive Transformation (FoPM) and JSPS Research Fellowship.
\end{acknowledgments}

\appendix

\section{Expression of multipoles in spinful $s$-$p$ hybridized system}
\label{sec: Appendix1}

In this Appendix, we show the definition of multipole operators used in the main text. 
For that purpose, we consider the atomic wave functions for $s$ and $p$ orbitals, whose angle dependence is given by 
\begin{align}
\cr&
\phi_{0}=\frac{1}{\sqrt{4\pi}}, 
\cr&
\phi_{p_x}=\sqrt{\frac{3}{4\pi}}\frac{x}{r} ,
\quad
\phi_{p_y}=\sqrt{\frac{3}{4\pi}}\frac{y}{r} ,
\quad
\phi_{p_z}=\sqrt{\frac{3}{4\pi}}\frac{z}{r}.
\quad
\end{align}

For the basis function $(\phi_0,\phi_{p_x},\phi_{p_y},\phi_{p_z})$, there are 16 multipole degrees of freedom in spinless space, which are given by~\cite{hayami2018microscopic} 
\begin{align}
&
Q_{0s}=
\left(
\begin{array}{c|ccc}
 1 & 0 & 0 & 0 \\ \hline
 0 & 0 & 0 & 0 \\
 0 & 0 & 0 & 0 \\
 0 & 0 & 0 & 0 \\
\end{array}
\right),
\ \
Q_{0p}=
\left(
\begin{array}{c|ccc}
 0 & 0 & 0 & 0 \\ \hline 
 0 & 1 & 0 & 0 \\
 0 & 0 & 1 & 0 \\
 0 & 0 & 0 & 1 \\
\end{array}
\right),
\cr &
Q_{x}=\frac{1}{\sqrt{3}}
\left(
\begin{array}{c|ccc}
 0 & 1 & 0 & 0 \\  \hline
 1 & 0 & 0 & 0 \\
 0 & 0 & 0 & 0 \\
 0 & 0 & 0 & 0 \\
\end{array}
\right),
\ \
Q_{y}=\frac{1}{\sqrt{3}}
\left(
\begin{array}{c|ccc}
 0 & 0 & 1 & 0 \\  \hline
 0 & 0 & 0 & 0 \\
 1 & 0 & 0 & 0 \\
 0 & 0 & 0 & 0 \\
\end{array}
\right),
\cr &
Q_{z}=\frac{1}{\sqrt{3}}
\left(
\begin{array}{c|ccc}
 0 & 0 & 0 & 1 \\  \hline
 0 & 0 & 0 & 0 \\
 0 & 0 & 0 & 0 \\
 1 & 0 & 0 & 0 \\
\end{array}
\right),
\ \
\bm{Q}=(Q_{x},Q_{y},Q_{z}),
\cr &
M_{x}=
\left(
\begin{array}{c|ccc}
 0 & 0 & 0 & 0 \\  \hline
 0 & 0 & 0 & 0 \\
 0 & 0 & 0 & -i \\
 0 & 0 & i & 0 \\
\end{array}
\right),
\ \
M_{y}=
\left(
\begin{array}{c|ccc}
 0 & 0 & 0 & 0 \\  \hline
 0 & 0 & 0 & i \\
 0 & 0 & 0 & 0 \\
 0 & -i & 0 & 0 \\
\end{array}
\right),
\cr &
M_{z}=
\left(
\begin{array}{c|ccc}
 0 & 0 & 0 & 0 \\  \hline
 0 & 0 & -i & 0 \\
 0 & i & 0 & 0 \\
 0 & 0 & 0 & 0 \\
\end{array}
\right),
\ \
\bm{M}=(M_{x},M_{y},M_{z}),
\cr &
T_{x}=\frac{1}{3\sqrt{3}}
\left(
\begin{array}{c|ccc}
 0 & i & 0 & 0 \\  \hline
 -i & 0 & 0 & 0 \\
 0 & 0 & 0 & 0 \\
 0 & 0 & 0 & 0 \\
\end{array}
\right),
\ \
T_{y}=\frac{1}{3\sqrt{3}}
\left(
\begin{array}{c|ccc}
 0 & 0 & i & 0 \\  \hline
 0 & 0 & 0 & 0 \\
 -i & 0 & 0 & 0 \\
 0 & 0 & 0 & 0 \\
\end{array}
\right),
\cr &
T_{z}=\frac{1}{3\sqrt{3}}
\left(
\begin{array}{c|ccc}
 0 & 0 & 0 & i \\  \hline
 0 & 0 & 0 & 0 \\
 0 & 0 & 0 & 0 \\
 -i & 0 & 0 & 0 \\
\end{array}
\right),
\ \
\bm{T}=(T_{x},T_{y},T_{z}),
\cr &
Q_{u}=\frac{1}{5}
\left(
\begin{array}{c|ccc}
 0 & 0 & 0 & 0 \\  \hline
 0 & -1 & 0 & 0 \\
 0 & 0 & -1 & 0 \\
 0 & 0 & 0 & 2 \\
\end{array}
\right),
\ \
Q_{v}=\frac{\sqrt{3}}{5}
\left(
\begin{array}{c|ccc}
 0 & 0 & 0 & 0 \\  \hline
 0 & 1 & 0 & 0 \\
 0 & 0 & -1 & 0 \\
 0 & 0 & 0 & 0 \\
\end{array}
\right),
\cr &
Q_{yz}=\frac{\sqrt{3}}{5}
\left(
\begin{array}{c|ccc}
 0 & 0 & 0 & 0 \\  \hline
 0 & 0 & 0 & 0 \\
 0 & 0 & 0 & 1 \\
 0 & 0 & 1 & 0 \\
\end{array}
\right),
\ \
Q_{zx}=\frac{\sqrt{3}}{5}
\left(
\begin{array}{c|ccc}
 0 & 0 & 0 & 0 \\  \hline
 0 & 0 & 0 & 1 \\
 0 & 0 & 0 & 0 \\
 0 & 1 & 0 & 0 \\
\end{array}
\right),
\cr &
Q_{xy}=\frac{\sqrt{3}}{5}
\left(
\begin{array}{c|ccc}
 0 & 0 & 0 & 0 \\  \hline
 0 & 0 & 1 & 0 \\
 0 & 1 & 0 & 0 \\
 0 & 0 & 0 & 0 \\
\end{array}
\right).
\end{align}
By using the addition rule between the above spinless multipoles and the Pauli matrix $\bm{\sigma}=(\sigma_x, \sigma_y, \sigma_z)$, the 48 multipoles in spinful space are obtained in Table~\ref{tab:mp}~\cite{kusunose2020complete}.

\begin{table*}[t!]
\caption{
Active spin-dependent dimensionless multipoles in the spinful $s$-$p$ Hilbert space in the form, $c_{x}\sigma_{x}+c_{y}\sigma_{y}+c_{z}\sigma_{z}$. 
We have introduced the abbreviations, $Q_{u}^{(\pm)}=Q_{u}\pm\sqrt{3}Q_{v}$, $Q_{v}^{(\pm)}=\pm\sqrt{3}Q_{u}+Q_{v}$, for notational simplicity.
}
\label{tab:mp}
\centering
\renewcommand{\arraystretch}{2.0}
\begin{tabular}{ccccccccc} \hline\hline
rank & E & $(c_{x},c_{y},c_{z})$ & ET & $(c_{x},c_{y},c_{z})$ & M & $(c_{x},c_{y},c_{z})$ & MT & $(c_{x},c_{y},c_{z})$ \\ \hline
0 & $Q^{\rm (s)}_0$ & $\displaystyle \frac{1}{\sqrt{3}}(M_{x},M_{y},M_{z})$ & $G^{\rm (s)}_0$ & $\displaystyle \frac{1}{\sqrt{3}}(T_{x},T_{y},T_{z})$ & $M^{\rm (s)}_0$ & $\displaystyle \frac{1}{\sqrt{3}}(Q_{x},Q_{y},Q_{z})$ \\ \hline
1 & $Q_{x}^{\rm (s)}$ & $\displaystyle \frac{1}{\sqrt{2}}(0,T_{z},-T_{y})$ & $G_{x}^{\rm (s)}$ & $\displaystyle \frac{1}{\sqrt{2}} (0,M_{z},-M_{y})$ & $M_{s,x}^{(\rm s)}$ & $(Q_{0s},0,0)$ & $T_{x}^{\rm (s)}$ & $\displaystyle \frac{1}{\sqrt{2}} (0,Q_{z},-Q_{y})$ \\
& $Q_{y}^{\rm (s)}$ & $\displaystyle \frac{1}{\sqrt{2}} (-T_{z},0,T_{x})$ & $G_{y}^{\rm (s)}$ & $\displaystyle \frac{1}{\sqrt{2}} (-M_{z},0,M_{x})$ & $M_{s,y}^{(\rm s)}$ & $(0,Q_{0s},0)$ & $T_{y}^{\rm (s)}$ & $\displaystyle \frac{1}{\sqrt{2}} (-Q_{z},0,Q_{x})$ \\
& $Q_{z}^{\rm (s)}$ & $\displaystyle \frac{1}{\sqrt{2}} (T_{y},-T_{x},0)$ & $G_{z}^{\rm (s)}$ & $\displaystyle \frac{1}{\sqrt{2}} (M_{y},-M_{x},0)$ & $M_{s,z}^{(\rm s)}$ & $(0,0,Q_{0s})$ & $T_{z}^{\rm (s)}$ & $\displaystyle \frac{1}{\sqrt{2}} (Q_{y},-Q_{x},0)$ \\
& & & & & $M^{(\rm s)}_{p,x}$ & $(Q_{0p},0,0)$ \\
& & & & & $M^{(\rm s)}_{p,y}$ & $(0,Q_{0p},0)$ \\
& & & & & $M^{(\rm s)}_{p,z}$ & $(0,0,Q_{0p})$ \\
& & & & & $M^{(\rm s)}_{a,x}$ & $\displaystyle \sqrt{\frac{3}{10}} \left(-\frac{1}{\sqrt{3}}Q_u^{(-)},Q_{xy},Q_{zx}\right)$ \\
& & & & & $M^{(\rm s)}_{a,y}$ & $\displaystyle \sqrt{\frac{3}{10}} \left(Q_{xy},-\frac{1}{\sqrt{3}}Q_u^{(+)},Q_{yz} \right)$ \\
& & & & & $M^{(\rm s)}_{a,z}$ & $\displaystyle \sqrt{\frac{3}{10}} \left(Q_{zx},Q_{yz},\frac{2}{\sqrt{3}}Q_u \right)$ \\ \hline
2 & $Q^{\rm (s)}_u$ & $\displaystyle \frac{1}{\sqrt{6}}(-M_{x},-M_{y},2M_z)$ & $G^{\rm (s)}_u$ & $\displaystyle \frac{1}{\sqrt{6}}(-T_{x},-T_{y},2T_z)$ & $M^{\rm (s)}_u$ & $\displaystyle \frac{1}{\sqrt{6}}(-Q_{x},-Q_{y},2Q_z)$ & $T^{\rm (s)}_u$ & $\displaystyle \frac{1}{\sqrt{2}}( Q_{yz},- Q_{zx},0)$ \\
& $Q^{\rm (s)}_v$ & $\displaystyle \frac{1}{\sqrt{2}} (M_x,-M_y,0)$ & $G^{\rm (s)}_v$ & $\displaystyle \frac{1}{\sqrt{2}} (T_x,-T_y,0)$ & $M^{\rm (s)}_v$ & $\displaystyle \frac{1}{\sqrt{2}}(Q_x,-Q_y,0)$ & $T^{\rm (s)}_v$ & $\displaystyle \frac{1}{\sqrt{6}} (Q_{yz},Q_{zx}, -2 Q_{xy})$ \\
& $Q^{\rm (s)}_{yz}$ & $\displaystyle \frac{1}{\sqrt{2}} (0,M_{z},M_y)$ & $G^{\rm (s)}_{yz}$ & $\displaystyle \frac{1}{\sqrt{2}} (0,T_{z},T_y)$ & $M^{\rm (s)}_{yz}$ & $\displaystyle \frac{1}{\sqrt{2}}(0,Q_{z},Q_y)$ & $T^{\rm (s)}_{yz}$ & $\displaystyle \frac{1}{\sqrt{6}} (-Q_v^{(+)},-Q_{xy},Q_{zx})$ \\
& $Q^{\rm (s)}_{zx}$ & $\displaystyle \frac{1}{\sqrt{2}} (M_z,0,M_x)$ & $G^{\rm (s)}_{zx}$ & $\displaystyle \frac{1}{\sqrt{2}} (T_z,0,T_x)$ & $M^{\rm (s)}_{zx}$ & $\displaystyle \frac{1}{\sqrt{2}}(Q_z,0,Q_x)$ & $T^{\rm (s)}_{zx}$ & $\displaystyle \frac{1}{\sqrt{6}} (Q_{xy},-Q_v^{(-)},-Q_{yz})$ \\
& $Q^{\rm (s)}_{xy}$ & $\displaystyle \frac{1}{\sqrt{2}}(M_{y},M_x,0)$ & $G^{\rm (s)}_{xy}$ & $\displaystyle \frac{1}{\sqrt{2}} (T_{y},T_x,0)$ & $M^{\rm (s)}_{xy}$ & $\displaystyle \frac{1}{\sqrt{2}}(Q_{y},Q_x,0)$ & $T^{\rm (s)}_{xy}$ & $\displaystyle \frac{1}{\sqrt{6}} (-Q_{zx},Q_{yz},2Q_{v})$ \\ \hline
3 & & & & & $M^{\rm (s)}_{xyz}$ & $\displaystyle \frac{1}{\sqrt{3}}(Q_{yz},Q_{zx},Q_{xy})$ \\
& & & & & $M^{\alpha {\rm (s)}}_{x}$ & $\displaystyle -\frac{1}{\sqrt{5}}\left(\frac{\sqrt{3}}{2}Q_u^{(-)},Q_{xy},Q_{zx} \right)$ \\
& & & & & $M^{\alpha {\rm (s)}}_{y}$ & $\displaystyle -\frac{1}{\sqrt{5}}\left(Q_{xy},\frac{\sqrt{3}}{2}Q_u^{(+)},Q_{yz}\right)$ \\
& & & & & $M^{\alpha {\rm (s)}}_{z}$ & $\displaystyle -\frac{1}{\sqrt{5}}(Q_{zx},Q_{yz},-\sqrt{3}Q_{u})$ \\
& & & & & $M^{\beta {\rm (s)}}_{x}$ & $\displaystyle \frac{1}{\sqrt{3}} \left(-\frac{1}{2}Q_v^{(+)},Q_{xy},-Q_{zx}\right)$ \\
& & & & & $M^{\beta {\rm (s)}}_{y}$ & $\displaystyle \frac{1}{\sqrt{3}} \left(-Q_{xy},-\frac{1}{2}Q_v^{(-)},Q_{yz} \right)$ \\
& & & & & $M^{\beta {\rm (s)}}_{z}$ & $\displaystyle \frac{1}{\sqrt{3}} (Q_{zx},-Q_{yz},Q_{v})$ \\
\hline\hline
\end{tabular}
\end{table*}

\section{Derivation of interaction between matter and electromagnetic wave}
\label{sec:coupling}

Let us begin with the weak relativistic Hamiltonian with the spin-orbit coupling,
\begin{align}
&
H=\frac{\bm{\pi}^{2}}{2m}-e\phi
+H_{\rm Z}+H_{\rm SOC}+H_{\rm D},
\cr&\quad
H_{\rm Z}=\mu_{\rm B}\bm{\sigma}\cdot\bm{B},
\cr&\quad
H_{\rm SOC}=
-\frac{e\hbar}{8m^{2}c^{2}}[(\bm{\sigma}\times\bm{\pi})\cdot\bm{E}_{\rm tot}+\bm{E}_{\rm tot}\cdot(\bm{\sigma}\times\bm{\pi})],
\cr&\quad
H_{\rm D}=\frac{e\hbar^{2}}{8m^{2}c^{2}}(\bm{\nabla}\cdot\bm{E}_{\rm tot})
\end{align}
where $\bm{\pi}=\bm{p}+(e/c)\bm{A}$ is the gauge-invariant kinetic momentum operator with the electron charge $-e$, and $H_{\rm Z}$, $H_{\rm SOC}$, and $H_{\rm D}$ represent the Zeeman, spin-orbit-coupling, and Darwin terms, respectively.
$\mu_{\rm B}=e\hbar/2mc$ is the Bohr magneton.
The electric field consists of those from the static ionic potential to bound an electron, $\bm{E}_{\rm ion}(\bm{r})=-\bm{\nabla}\phi$ and an external field $\bm{E}(\bm{r},t)=-(1/c)\partial\bm{A}/\partial t$, i.e. $\bm{E}_{\rm tot}=\bm{E}_{\rm ion}+\bm{E}$.
We adopt the Coulomb (radiation) gauge, $\bm{\nabla}\cdot\bm{A}=0$.
Hereafter, we assume that the ionic potential is spherically symmetric, $\phi(\bm{r})=\phi(r)$ and the electromagnetic field is regarded as a classical field so that $\bm{E}$, $\bm{B}$, and $\bm{A}$ are commutable with each other.

The time-dependent Schr\"odinger equation is given by
\begin{align}
i\hbar\frac{\partial}{\partial t}\psi=H\psi.
\label{eq:se}
\end{align}
Considering a unitary (gauge) transformation $U$ to (\ref{eq:se}) from the left, we obtain the transformed one with the prime as
\begin{align}
&
i\hbar\frac{\partial}{\partial t}\psi'=H'\psi',
\cr&\quad
\psi'=U\psi,
\quad
H'=UHU^{-1}+i\hbar\frac{\partial U}{\partial t}U^{-1}.
\label{eq:gth}
\end{align}

Now, let us consider the case $U=e^{ie\bm{r}\cdot\bm{A}(\bm{r},t)/c\hbar}$, which is known as Power-Zienau-Woolley transformation (in the long wavelength limit)~\cite{Scully1997, Gerry2005, steck2023}.
Using the formula $[\bm{p},U]=-i\hbar\bm{\nabla}U$, we obtain
\begin{align}
U\bm{p}U^{-1}&=\bm{p}-[\bm{p},U]U^{-1}
=\bm{p}-\frac{e}{c}\bm{\nabla}(\bm{r}\cdot\bm{A})
\cr&
=\bm{p}-\frac{e}{c}[\bm{A}+\bm{r}\times(\bm{\nabla}\times\bm{A})+\bm{r}\cdot\bm{\nabla}\bm{A}]
\cr&
\simeq
\bm{p}-\frac{e}{c}\bm{A}(t)-\frac{e}{c}\bm{r}\times\bm{B}(t),
\end{align}
where we have neglected $\bm{r}$ dependences in $\bm{A}$ and $\bm{B}$ and the last term, $-(e/c)\bm{r}\cdot\bm{\nabla}\bm{A}$, in the last step, as $\bm{A}$ is slowly varying field as compared with the atomic scale.
Thus, we have
\begin{align}
U\bm{\pi}U^{-1}\simeq\bm{p}-\frac{e}{c}(\bm{r}\times\bm{B}(t)).
\end{align}
Moreover, the second term of $H'$ in (\ref{eq:gth}) is
\begin{align}
i\hbar\frac{\partial U}{\partial t}U^{-1}=-\frac{e}{c}\bm{r}\cdot\frac{\partial\bm{A}}{\partial t}
=e\bm{r}\cdot\bm{E}(t).
\end{align}
Using these relations, we obtain the transformed Hamiltonian as
\begin{align}
&
H'=H_{0}+H_{\rm int},
\cr&
H_{0}=\frac{\bm{p}^{2}}{2m}
-e\phi(r)
+\frac{\xi}{2}\bm{l}\cdot\bm{\sigma}
+\frac{e\hbar^{2}}{8m^{2}c^{2}}(\bm{\nabla}\cdot\bm{E}_{\rm ion}),
\cr&
H_{\rm int}=-\bm{\mu}_{\rm e}\cdot\bm{E}(t)-\bm{\mu}_{\rm m}\cdot\bm{B}(t)
+\frac{e^{2}}{2mc^{2}}(\bm{r}\times\bm{B}(t))^{2}
\cr&\quad
+\frac{e^{2}\hbar}{4m^{2}c^{3}}\sum_{\alpha\beta}m_{\alpha\beta}'E_{\alpha}(t)B_{\beta}(t),
\end{align}
where $\xi=e\hbar^{2}(d\phi/dr)/(2m^{2}c^{2}r)$.
The third term of $H_{\rm int}$ represents the diamagnetic contribution.
We have introduced the electric and magnetic dipoles as
\begin{align}
&
\bm{\mu}_{\rm e}=-e\bm{r}+\bm{\mu}_{\rm e}',
\cr&
\bm{\mu}_{\rm m}=-\mu_{\rm B}(\bm{l}+\bm{\sigma})+\bm{\mu}_{\rm m}',
\end{align}
where $\bm{l}=(\bm{r}\times\bm{p})/\hbar$ is the dimensionless orbital angular momentum operator, and $m_{\alpha\beta}'=r_{\alpha}\sigma_{\beta}-3(\bm{r}\cdot\bm{\sigma})\delta_{\alpha\beta}+2(\bm{r}\cdot\bm{\sigma})\delta_{\alpha\beta}$ is a combination of magnetic quadrupole and monopole operators.
Here, the spin-orbit corrections are given by
\begin{align}
&
\bm{\mu}_{\rm e}'=\frac{e\hbar}{4m^{2}c^{2}}(\bm{\sigma}\times\bm{p}),
\\&
\bm{\mu}_{\rm m}'=-\frac{e\xi}{2c\hbar}(\bm{m}_{\rm a}+r^{2}\bm{\sigma}),
\end{align}
where $\bm{m}_{\rm a}=(\bm{r}\cdot\bm{\sigma})\bm{r}-3r^{2}\bm{\sigma}$ is an anisotropic magnetic dipole operator, and the second term in $\bm{\mu}_{\rm m}'$ is the correction to the spin dipole.

The strength of the SOC in the main text is given by the radial average $\lambda=\braket{\xi}_{pp}$ with respect to the radial $p$ wave function.

\section{Effect of spin-orbit correction in interaction term}
\label{sec: Appendix2}

We discuss the effect of the spin-orbit correction of $\bm{\mu}_{\rm e}'$ in Eq.~(\ref{eq:Ht}).
The interaction by taking account of this term is expressed as
\begin{align}
\label{eq:Ht2}
H_{\rm int}'(t) =  -\bm{\mu}_{\rm e}'\cdot\bm{E}(t) -\bm{\mu}_{\rm m}\cdot\bm{B}(t),
\quad
\bm{\mu}_{\rm m}\simeq-\mu_{\rm B}(\bm{l}+\bm{\sigma}),
\end{align}
where $\bm{\mu}_{\rm e}'=\gamma \bm{Q}^{\rm (s)}$ with $\gamma=3\sqrt{2}e\hbar^{2}\braket{r^{-1}}_{sp}/(4m^{2}c^{2})$, and the electric-field strength is given by $E'=\gamma E_{0}$.

By performing a similar procedure in Sec.~\ref{sec: Effective Hamiltonian}, we derive an effective Hamiltonian up to $\Omega^{-2}$, which is given by
\begin{align} 
\label{eq:Ham_app}
H'_\mathrm{eff} =& H_0 + H'_1+H'_2+O(\Omega^{-2}).
\end{align}
The first-order term is given by 
\begin{align}
&
H'_1=-\frac{ \sin 2 \eta}{4\Omega} (H'^{\perp}_1 + H'^{\parallel}_1),
\\&\quad
H'^{\perp}_1=   \frac{E'B}{\sqrt{6}} \left(  4G_0^{\rm (s)} +\sqrt{2} G_u^{\rm (s)} 
\right),
\\&\quad
H'^{\parallel}_1= \frac{1}{54} \left(E'^2+54 B^2\right)( M_z+2 M^{\rm (s)}_{s,z} )
\cr&\quad
 +\frac{1}{162} [2\left(E'^2+162 B^2\right)  M^{\rm (s)}_{p,z}
 +5\sqrt{10}  E'^2   M^{\rm (s)}_{a,z}]. \quad
\end{align}
In contrast to the case without the spin-orbit correction in the main text, this correction gives rise to the coupling to the ET monopole in the first order.

The second-order term is given by 
\begin{align}
&
H'_2=\frac{1}{8\Omega^2} [(H'^{\perp}_{2a}+H'^{\parallel}_{2a})
+ \cos 2 \eta  (H'^{\perp}_{2b}+H'^{\parallel}_{2b})],
\\&\quad
H'^{\perp}_{2a}=\frac{E' B  \lambda}{\sqrt{2}}  T_z -\frac{E' B}{2}   (2 \Delta-\lambda) T^{\rm (s)}_z,
\\&\quad
H'^{\parallel}_{2a}= \frac{E'^2}{162} (2 \Delta-\lambda) (12Q_{0s}-4Q_{0p}-5Q_u)
\cr&\hspace{1.2cm}
+\frac{1}{27 \sqrt{3}}[2 E'^2 \Delta - 3 \left(E'^2+18 B^2\right)\lambda  ] Q^{\rm (s)}_0
\cr&\hspace{1.2cm}
- \frac{1}{27 \sqrt{6}}(2 E'^2 \Delta +27 B^2 \lambda) Q^{\rm (s)}_u,
\\&\quad
H'^{\perp}_{2b}=-\frac{E' B}{6}   (2 \Delta-3 \lambda)  M^{\rm (s)}_{xy},
\\&\quad
H'^{\parallel}_{2b}= \frac{5 E'^2 (2 \Delta-\lambda) }{54\sqrt{3}} Q_v
+\frac{ \left(2 \Delta E'^2 - 27 B^2 \lambda\right)}{27\sqrt{2}} Q^{\rm (s)}_v.
\cr&
\end{align}
Although the coefficients are different from those in the main text, the same type of multipoles are induced.

\bibliographystyle{apsrev}
\bibliography{ref.bib}
\end{document}